\documentclass[aps,prb,showpacs,twocolumn,reprint,superscriptaddress]{revtex4-2}
\usepackage{amsmath}
\usepackage{amssymb}
\usepackage{graphicx}
\usepackage{color}
\usepackage{multirow}
\usepackage{dcolumn}
\usepackage{textcomp}
\usepackage{longtable}
\usepackage{makecell}
\usepackage{soul, color, xcolor, ulem}

\begin{document}
	
	\title{Evolution of flat bands in MoSe$_2$/WSe$_2$ moiré lattices: A study combining machine learning and band unfolding methods}
	
	\author {Shengguo Yang}
	\affiliation{Key Laboratory for Matter Microstructure and Function of Hunan Province,
		Key Laboratory of Low-Dimensional Quantum Structures and Quantum Control of Ministry of Education, School of Physics and Electronics, Hunan Normal University, Changsha 410081, China}
	
	\author {Jiaxin Chen}
	\affiliation{Key Laboratory for Matter Microstructure and Function of Hunan Province,
		Key Laboratory of Low-Dimensional Quantum Structures and Quantum Control of Ministry of Education, School of Physics and Electronics, Hunan Normal University, Changsha 410081, China}
	
	\author {Chao-Fei Liu}
	\affiliation{School of Science, Jiangxi University of Science and Technology, Ganzhou 341000, China}
	
	\author {Mingxing Chen}
	\email{mxchen@hunnu.edu.cn}
	\affiliation{Key Laboratory for Matter Microstructure and Function of Hunan Province,
		Key Laboratory of Low-Dimensional Quantum Structures and Quantum Control of Ministry of Education, School of Physics and Electronics, Hunan Normal University, Changsha 410081, China}
	\affiliation{State Key Laboratory of Powder Metallurgy, Central South University,  Changsha 410083, China}
	
	\date{\today}
	
	\begin{abstract}
		Moiré lattices have served as the ideal quantum simulation platform for exploring novel physics due to the flat electronic bands resulting from the long wavelength moiré potentials. However, the large sizes of this type of system challenge the first-principles methods for full calculations of their electronic structures, thus bringing difficulties in understanding the nature and evolution of the flat bands. In this study, we investigate the electronic structures of moiré patterns of MoSe$_2$/WSe$_2$ by combining ab initio and machine learning methods. We find that a flat band with a bandwidth of about 5 meV emerges below the valence band edge at the K point for the H-stacking at a twist angle of 3.89$^{\circ}$ without spin-orbit coupling effect. Then, it shifts dramatically as the twist angle decreases and becomes about 20 meV higher than the valence band maximum for the twist angle of 3.15$^{\circ}$. Multiple ultra-flat bands emerge as the twist angle is reduced to 1.7$^{\circ}$. The spin-orbit coupling leads to a giant spin splitting comparable to that observed in the untwisted system (about 0.45 eV) and is nearly independent of twisting and stacking. As a result, the K-valley flat band remains the valence band maximum with the inclusion of spin-orbit coupling. Band unfolding reveals that the ultra-flat bands formed by the $\Gamma$ and K valleys show distinct behaviors. The $\Gamma$-valley flat bands are sensitive to the interlayer coupling, thus experiencing dramatic changes as the twist angle decreases. In contrast, the K-valley flat band, which shows a weak dependence on the interlayer coupling, is mainly modulated by structural reconstruction. Therefore, a relatively small angle (2.13$^{\circ}$) is required to generate the K-valley flat band, which experiences a transition from the honeycomb to the triangular lattice as the twist angle decreases. 
	\end{abstract}
	
	
	\maketitle
	\section{INTRODUCTION}
	Moiré lattices formed by twisting two-dimensional (2D) materials or aligning dissimilar ones have been the ideal platform for exploring novel physics such as strong-correlation effects~\cite{Cao2018a,Jiang2019,Kerelsky2019,Xie2019,Wang2020,Tang2020,Regan2020}, unconventional superconductivity~\cite{Cao2018,Yankowitz2019,Hao2021}, and topological phases~\cite{Jiao2020,Falson2020,Serlin2020,Choi2021,Nuckolls2020,Chen2019,Chen2021,Polshyn2020,Shen2021}. The key feature of this type of systems is the flat electronic bands due to the long wavelength periodic potentials, which is tunable by the twist angle. This unusual property was early discovered in twisted bilayer graphene (tBLG),  for which the flat bands emerge only at a few so-called magic angles~\cite{Bistritzer2011,Tarnopolsky2019a,Cao2018,Cao2018a}.  Inspired by this discovery,  twisted systems of many other 2D materials have been extensively studied~\cite{Xie2022twist,Zhao2020,Sun2022labr2,An2021moire,Kennes2020one,Li2024d,Venkateswarlu2020,Carr2018,Kang2013electronic,Guo2020shedding,Waters2020flat,Enaldiev2020,Kang2024evidence,Wang2024fractional,Anderson2023programming,Morales2021metal,Shen2021exotic,Claassen2022ultra}.  Interestingly,  the angles for generating the flat bands in these materials can be much larger than the magic angles for tBLG. For instance, for twisted bilayer transition metal dichalcogenides (tbTMDs), density-functional theory (DFT) calculations predicted that flat bands could be formed at an angle as large as 7$^{\circ}$~\cite{Wu2018,Wu2019,Xu2022,Naik2018,Naik2020,Wang2020,Venkateswarlu2020,Zhang2020,Zhang2021,Angeli2021,Xian2021,Devakul2021}. 
	
	The tbTMDs can exhibit much richer geometric and electronic structures than tBLG~\cite{Kennes2021a,Magorrian2021}. For instance, they can be obtained by twisting the TMD bilayers either in R-stacking or in H-stacking. Namely, their structure is dependent on the relative orientation of the two layers. As a result,  they show interesting geometric patterns such as honeycomb and triangular domain wall networks depending on the stacking order. In addition, the TMD bilayers possess two different groups of valleys, i.e., the $\Gamma$ and K valleys. 
		The flat bands originating from these valleys allow for exploring various physical models. For example, the $\Gamma$-valley flat bands have been proposed to realize the honeycomb and kagome lattice models~\cite{Angeli2021}, while the K-valley flat bands have been used to study the triangular Hubbard model~\cite{Wu2018, Pan2020quantum, Pan2020} and the Kane-Mele model~\cite{Wu2019}. 
	Moreover, the two valleys are energetically close. This feature along with the stacking order and twist angle can give rise to complicated electronic structures for tbTMDs. For instance, for the twisted bilayer of WSe$_2$ with an angle of 5.1$^{\circ}$ (a twist away from R-stacking) only the $\Gamma$-valley flat bands were observed~\cite{Pei2022,Pei2023}. As the twist angle decreases to 3$^{\circ}$, both the $\Gamma$- and K-valley flat bands were observed~\cite{Zhang2020}. For the twists near H-stacking, $\Gamma$-valley flat bands were observed for WSe$_2$ at angles of 57.4$^{\circ}$ and 57.5$^{\circ}$~\cite{Gatti2023,Zhang2020}. When the angle is slightly increased to 57.72$^{\circ}$, the K-valley flat band appears and becomes 129 meV higher than the $\Gamma$-valley one~\cite{Kundu2022}. 
	
    Among the tbTMDs, twisted heterobilayers, e.g., MoS$_2$/WS$_2$ and MoSe$_2$/WSe$_2$, have emerged as prominent candidates for investigating moiré interlayer excitons~\cite{sung2020broken,jin2019observation,alexeev2019resonantly,bai2020excitons,seyler2019signatures,tran2019evidence,brotons2020spin,baek2020highly,wang2021moire,zhang2018moire}, which can modulated by the moiré potential. Recently, a scanning tunneling microscope (STM) experiment shows the H-stacking MoSe$_2$/WSe$_2$ moiré system has been reported to exhibit a deep moiré potential exceeding 300 meV for the valence band as the moiré patterns reaching 13 nm. The moiré potential shows a non-monotonic behavior as a function of the moiré wavelength. As the moiré wavelength increases from 6 nm ($\sim$3.15$^\circ$) to 17 nm ($\sim$1.08$^\circ$), a signature of the moiré flat band emerges from the HMX site~\cite{Shabani2021}. However, its valley origin remains to be explored.
	
	DFT calculations play an important role in understanding the effects of twisting on the electronic structure of the twisted systems. Although DFT calculations were already performed for tbTMDs with a twist angle as small as 1.54$^{\circ}$~\cite{Naik2020}, they are extremely time-consuming since this type of lattice usually possesses a large number of atoms (the tbTMDs with such a twist angle contain 8322 atoms). Recently, deep-learning-based methods such as deep potential molecular dynamics (DeePMD)~\cite{Zhang2018,Wang2018,Zeng2023} and deep-learning DFT Hamiltonian (DeepH)~\cite{Li2022a,Gong2023} have demonstrated the ability to model large-scale systems with high efficiency and accuracy. Applying these methods to the twisted systems will be good for understanding the twist effect on the electronic properties of 2D materials. However, there are band unfoldings in the band structure due to the use of large supercells in the calculations, which hide the nature of the flat bands in these systems. In particular, for tbTMDs the $\Gamma$- and K-valley flat bands may be mixed up in the band structure from the supercell calculations, thus bringing difficulties in understanding the origin of the flat bands.	
	
	In this work, we investigate the geometric and electronic structures of MoSe$_2$/WSe$_2$ moiré lattices with twist angles in the range of 21.79$^{\circ}$ and 1.7$^\circ$ by combining the deep learning-based methods and a band unfolding technique. This strategy allows for identifying which valley the flat bands belong to and how they respond to the twist. This paper is organized as follows. We present the methods and the details of the calculations in Sec.~\ref{sec:METHODS}. Then, we show the benchmark tests of the deep potential (DP) and DeepH models in  Sec.~\ref{subsec:DP and DeepH models}, which is followed by discussions of the geometric properties of MoSe$_2$/WSe$_2$ moiré lattices presented in Sec.~\ref{subsec:Geometric structures} and the electronic structure without the inclusion of spin-orbit coupling (SOC) in Sec.~\ref{subsec:Electronic bands}. The effect of SOC on the band structure is discussed in Sec.~\ref{subsec:Effect of spin-orbit coupling}. Finally, the results are summarized in Sec.~\ref{sec:CONCLUSIONS}.
	
	\section{METHODS}
	\label{sec:METHODS}
\begin{figure}[!t]
		\includegraphics[width=1\linewidth]{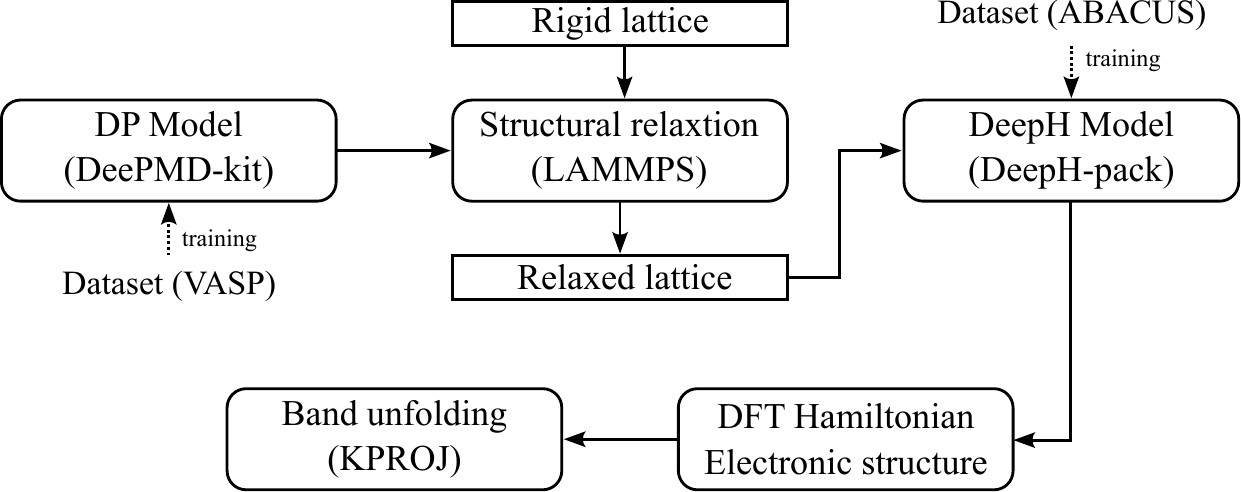}
		\caption{
                Workflow of computational scheme.
            }
		\label{workflow}
	\end{figure}

 We use a DP generated by the DeePMD-kit package to model the interatomic interactions for the MoSe$_2$/WSe$_2$ moiré lattices. The training data were obtained by DFT calculations using the Vienna Ab-initio Simulation Package (VASP)~\cite{Kresse1996}. The projector augmented wave (PAW) method was used to construct pseudopotentials~\cite{Kresse1999}. The energy cutoff for plane waves was 270 eV and the electronic exchange-correlation functional was parametrized by the formalism proposed by Perdew, Burke, and Ernzerhof (PBE)  within the generalized gradient approximation~\cite{Perdew1996}. The 2D Brillouin zone (BZ) was sampled using a 5$\times$5 $k$-mesh. The van der Waals forces between layers were corrected using the DFT-D3 method~\cite{Grimme2010}. Then, the LAMMPS program incorporating the DP model was used for structural relaxations~\cite{Plimpton1995,Bitzek2006,Thompson2022}, for which the threshold force is 10$^{-6}$ eV/\AA. The band structures were obtained using the DeepH-pack package. The atomic-orbital-based $ab$ $initio$ computation program at UStc (ABACUS)~\cite{ChenM2010,Li2016} was used to generate the training data for DeepH. The pseudo-atomic orbitals used in the ABACUS calculations are Se-$2s2p1d$, Mo-$4s2p2d1f$, and W-$4s2p2d2f$ with a radius cutoff of 10 a.u. and an energy cutoff of 100 Ry. Unfolded bands were obtained using the KPROJ program~\cite{Chen2018}, which projects the wave function of the moiré lattices onto the $k$-points in the BZ of the primitive cell.  Figure~\ref{workflow} shows the workflow of our computational scheme.

	\section{RESULTS AND DISCUSSIONS}
	\subsection{DP and DeepH models}
	\label{subsec:DP and DeepH models}
\begin{figure}[!t]
		\includegraphics[width=1\linewidth]{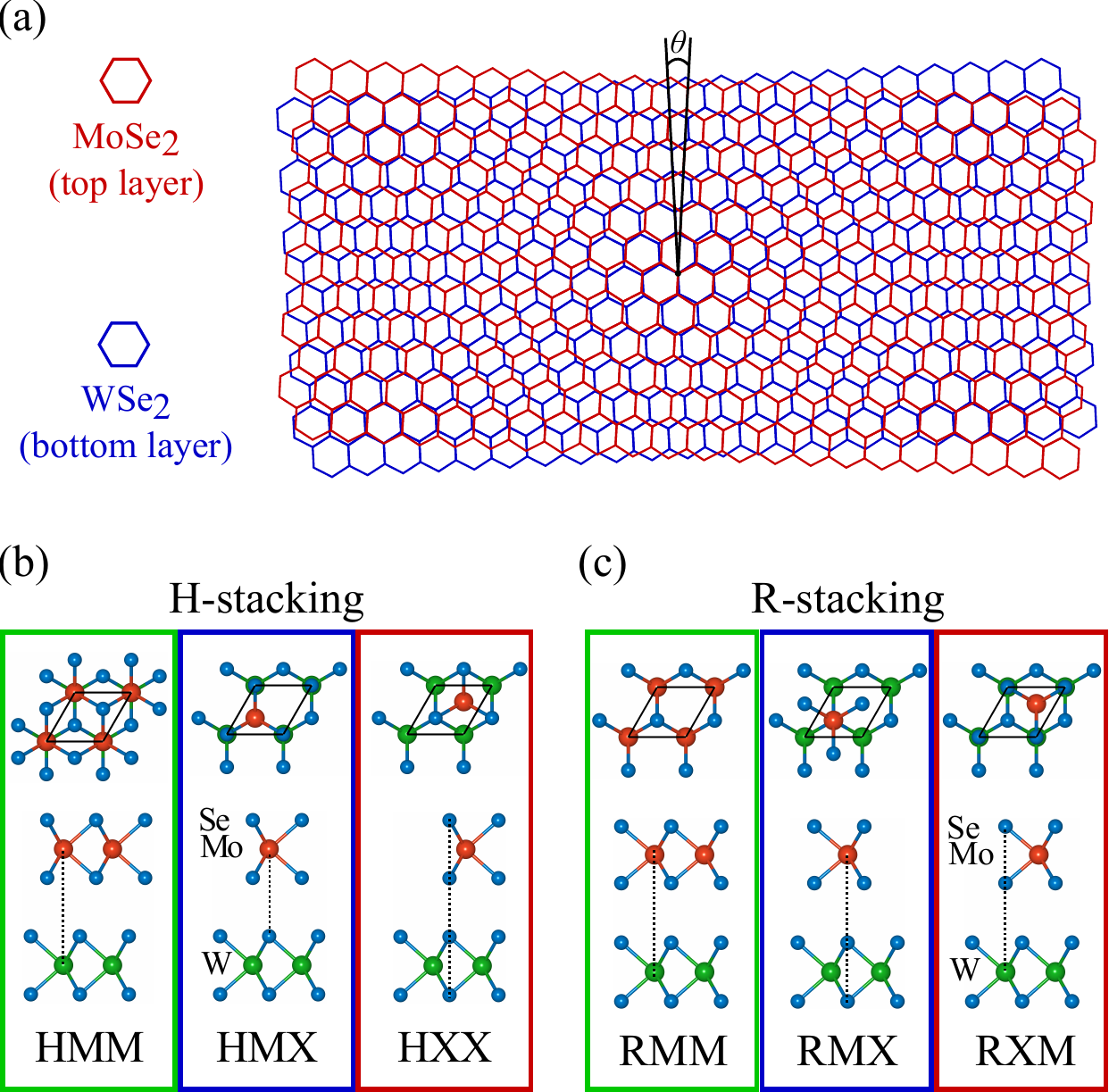}
		\caption{
			Geometrical structure of MoSe$_{2}$/WSe$_{2}$ moiré lattices. 
			(a) Schematic of a twisted bilayer of MoSe$_{2}$/WSe$_{2}$ with angle $\theta$.
			(b, c) Top and side views of high-symmetry stacking orders for the H-stacking and R-stacking.
		} 
		\label{stru_H_R}
	\end{figure}

	There are two types of configurations for a tbTMD. One has a twist angle $\theta$ away from the R-stacking, and the other is away from the H-stacking. Hereafter, they are named $R$-$\theta$ and $H$-$\theta$, respectively. There are three high-symmetry stacking orders for the untwisted bilayers in each kind of stacking. For H-stacking, they are named HMM, HMX, and HXX [Fig.~\ref{stru_H_R}(b)], respectively, where M represents the transition metal atoms Mo/W and X denotes the chalcogen atoms. Likewise, they are referred to as RMM, RMX, and RXM for R-stacking [Fig.~\ref{stru_H_R}(c)], respectively. The lattice constant used in our calculations is 3.282~\AA~for the 1 $\times$ 1 primitive cell. 
	
\begin{figure*}[!t]
		\includegraphics[width=1\linewidth]{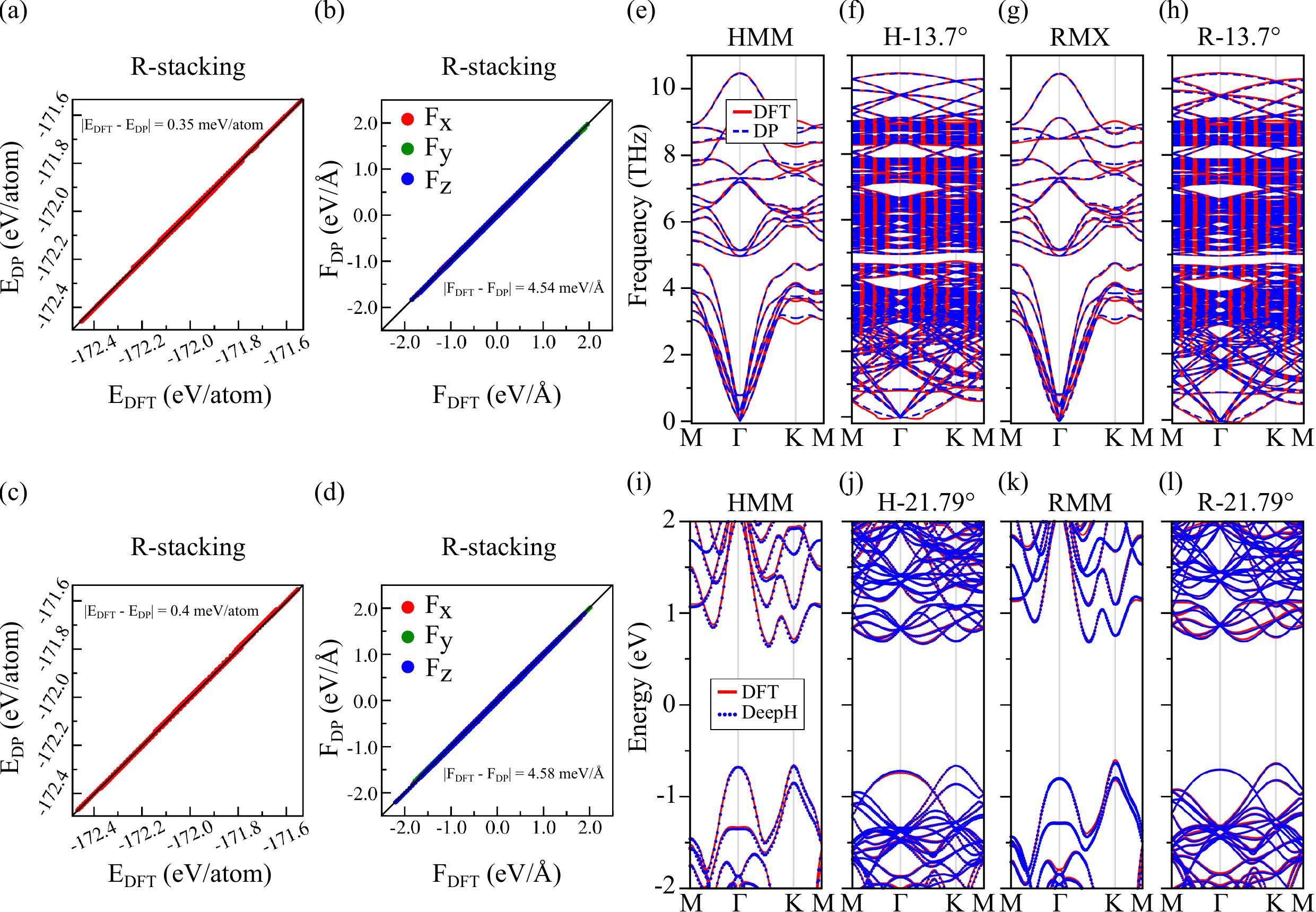}
		\caption{
			Benchmark tests of machine-learning models.
			Comparison of energies and atomic forces from the DP models and DFT calculations for $R$-stacking (a, b) and H-stacking (c, d).
			The MAE of energy and atomic force for the DP models is marked in the figures as black text.
			(e)-(h) Phonon spectrums for HMM, $H$-13.7$^{\circ}$, RMX, and $R$-13.7$^{\circ}$, respectively.
			(i)-(l) Band structures for HMM, $H$-21.79$^{\circ}$, RMM, and $R$-21.79$^{\circ}$, respectively.
			DFT results are also shown for comparison.
		} 
		\label{dp}
	\end{figure*}

	For the DP model, the dataset for training was generated by $NVT$ $ab$ $initio$ molecular dynamic (AIMD) simulations using VASP for a few structures, which include 2 $\times$ 2 supercells of the high-symmetry stacking orders shown in Fig.~\ref{stru_H_R}, those obtained by a relative shifting of the top layer in the stacking orders, and moiré structures with an angle of 21.79$^{\circ}$, i.e., $H$-$21.79$ and $R$-$21.79$. In addition, these structures under biaxial strains in the range from -2.5\% to 2.5\% were also included. For the DeepH model, structures for generating the dataset are 3 $\times$ 3 supercells of the high-symmetry stacking orders, the shifted ones, and the twisted structures with an angle of 13.17$^{\circ}$. The Hamiltonian hopping and overlapping matrix elements were obtained using ABACUS. Benchmark tests for the DP and DeepH models are shown in Fig.~\ref{dp}. It should be mentioned that we also calculated phonon dispersions for $H$-$13.7$ [Fig.~\ref{dp}(f)] and $R$-$13.7$ [Fig.~\ref{dp}(h)] using the DP model interfaced with LAMMPS and calculated electronic bands for $H$-$21.79$ [Fig.~\ref{dp}(j)] and $R$-$21.79$ [Fig.~\ref{dp}(l)] using DeepH, of which the structures were not included in the training. However, the results from the deep-learning models show good agreements with those from DFT calculations.    
In addition, we have also performed DeepH calculations for both $H$- and $R$-6.01$^\circ$, whose results are given in Appendix~\ref{sec:deeph model} (see Fig.~\ref{band_abacus_deeph_H_R_6.01}). The band structures derived from DeepH is also in good agreement with those obtained by DFT. These results suggest the high precision of the models.

	\subsection{Geometric structures}
	\label{subsec:Geometric structures}
\begin{figure*}[t]
		\includegraphics[width=1\linewidth]{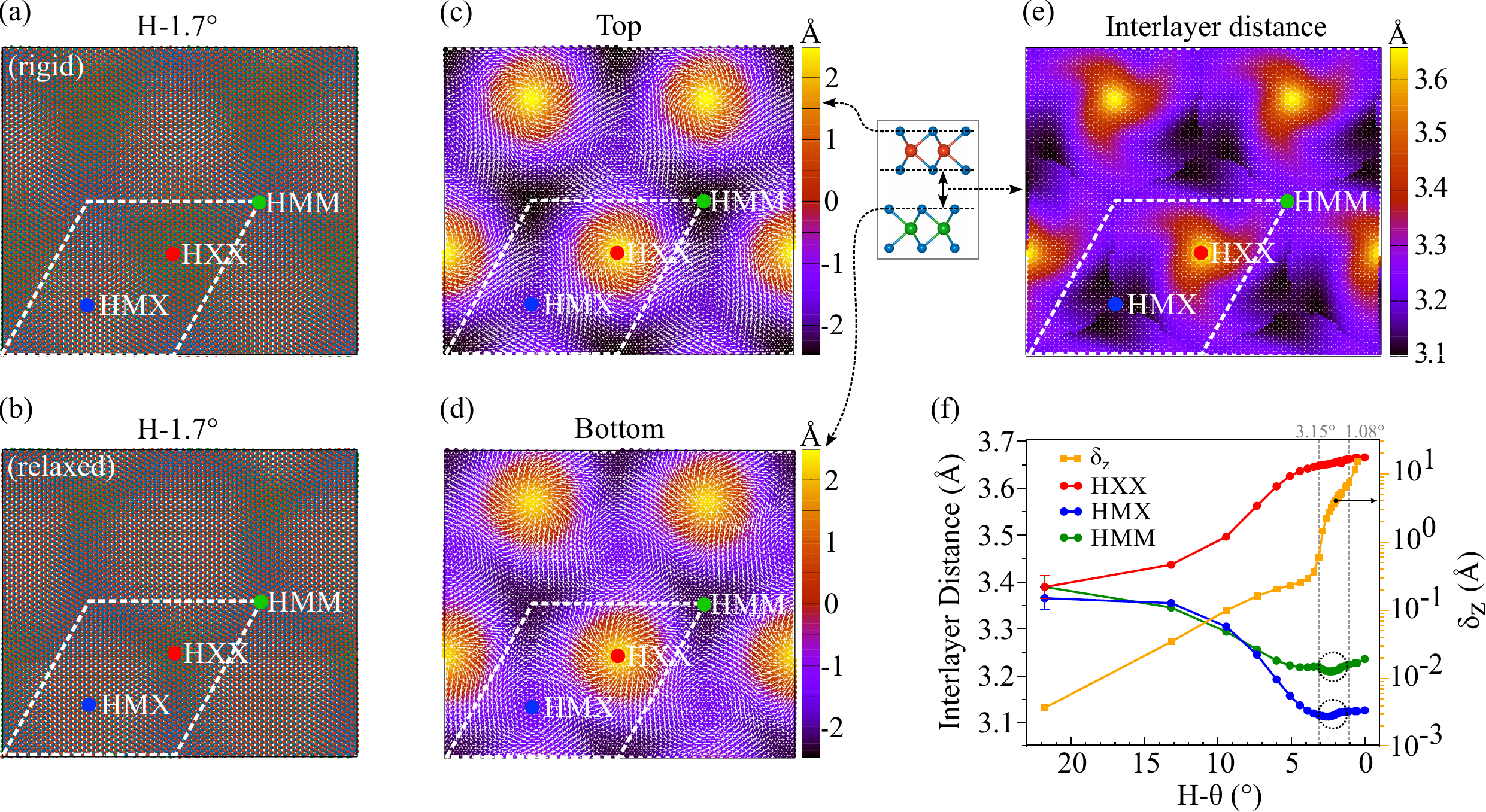}
		\caption{
			Geometrical structure of twisted MoSe$_{2}$/WSe$_{2}$ heterobilayers. 
			(a, b) Top view of the geometric structure for $H$-1.7$^{\circ}$ without and with lattice reconstruction.
			The parallelogram formed by the white dotted line represents a primitive cell of the moiré lattice.
			The blue, red, and green dots represent the three highly-symmetric stacking configurations, corresponding to HMX, HXX, and HMM in Fig. 1(b).  
			(c, d)~The relaxation induced in-plane and out-of-plane displacements for the Se atoms on the surface of MoSe$_2$ (top) and WSe$_2$ (bottom), respectively.
			Out-of-plane displacements are encoded as color bars, where positive/negative values indicate upward/downward displacements in relation to the reference point.
			The white arrows represent the in-plane displacements, for which, the maximum is about 0.37 and 0.39~\AA~in (c) and (d).
			(e)~The distribution of local interlayer distances for the relaxed structure in (b).
			(f)~Lattice reconstruction ($\delta_z$, the yellow rectangle-line), which is the difference between the maximum and minimum for the color bars in (c).
			The red, blue, and green dot lines denote the localized interlayer distance of three high-symmetry stacking sites.
		}
		\label{stru_H}
	\end{figure*}

	Figure~\ref{stru_H}(a) shows a rigid moiré structure for $H$-1.7$^{\circ}$, which exhibits a triangular pattern. 
	The pattern is then changed into a hexagonal domain-wall network after structural relaxation due to geometric reconstruction [see Fig.~\ref{stru_H}(b)]. The landscape of the reconstruction is in agreement with experimental results~\cite{McGilly2020,Weston2020,Rosenberger2020,Nieken2022,VanWinkle2023,Shabani2021,Rupp2023} and previous DFT calculations~\cite{Enaldiev2021}. Similar patterns were also seen for the twisted homobilayers~\cite{Edelberg2020a,Enaldiev2020}. Like previous studies~\cite{Geng2021,Carr2018,Li2021a,Zhang2021,Geng2022,Xie2023,Jia2024}, we also decompose the reconstruction into in-plane (IP) and out-of-plane (OOP) displacements. Figures~\ref{stru_H}(c) and \ref{stru_H}(d) show the displacements of the Se atoms on the surface of MoSe$_2$ and WSe$_2$, where the IP displacements are denoted by the white arrows and the OOP displacements are shown in colors.
	The IP displacements of the Se atoms on the surface of MoSe$_2$ show a clockwise rotation around the HMX site and a counterclockwise rotation around the HXX site [Fig.~\ref{stru_H}(c)]. The IP displacements reach the maximum (about 0.37~\AA~for $H$-1.7$^{\circ}$) in the region between the HMX and HXX sites. The chiral atomic displacements are also seen for the Se atoms on the surface of WSe$_2$ [Fig.~\ref{stru_H}(d)]. However, they are in opposite directions to those for MoSe$_2$. This trend is consistent with previous studies~\cite{Cantele2020,Yananose2021}. 
	
	For the OOP displacement, the surface Se atoms exhibit fluctuations up to about 5~\AA~and 4.5~\AA~for MoSe$_2$ and WSe$_2$, respectively. The fluctuations result in an apex in the HXX region and a basin at the HMM site. The interlayer distance has the largest value at HXX and the smallest value at HMX [Fig.~\ref{stru_H}(e)], which shows the same trend as the untwisted bilayers [see Fig.~\ref{layer_dis_band_H_R}(a) in Appendix~\ref{sec:untwisted_bilayer}]. The twisted structures near R-stacking show the same trend in the layer distance. We quantify the fluctuation in the OOP reconstruction by defining $\delta_z = OPP_{max}-OPP_{mini}$, which is shown in Fig.~\ref{stru_H}(f). It increases monotonically as the twist angle decreases from 21.79$^{\circ}$ to near 3.15$^{\circ}$. Then it increases quickly as the angle further decreases. This feature is related to different responses of the layer distances at the three high-symmetry sites to the twist. The layer distances in HMM and HMX regions roughly decrease as the twist angle decreases. In contrast, it increases in the HXX region [Fig.~\ref{stru_H}(f)]. 
	
\begin{figure*}[!t]
		\includegraphics[width=1\linewidth]{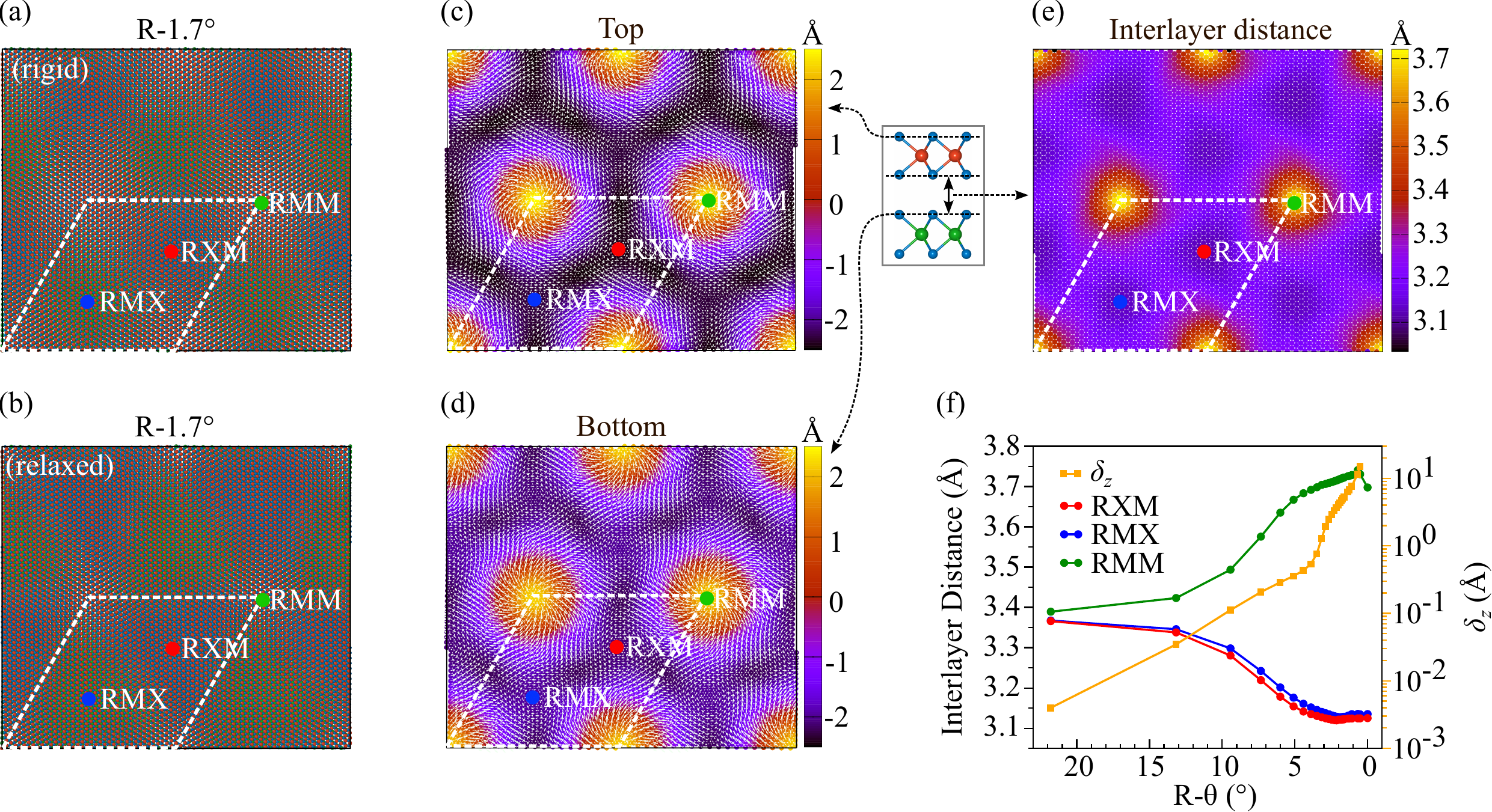}
		\caption{
			Geometrical structure of MoSe$_{2}$/WSe$_{2}$ R-stacking moiré lattice.
			(a, b)~Top view of the atomic structure of R-1.7$^{\circ}$ before and after relaxation.
			The white parallelogram denotes the primitive cell of the moiré lattice.
			(c, d)~Displacements of the Se atoms on the surface of MoSe$_2$ (top) and WSe$_2$ (bottom), respectively.
			(e)~Interlayer distances for the structure shown in (b).
			(f)~Lattice reconstruction ($\delta_z$) and the interlayer distances for the RXM, RMX, and RMM regions in the moiré lattice as functions of the twist angle.
		} 
		\label{stru_R}
	\end{figure*}

		The results for the R-$\theta$ moiré lattices show a similar trend in the lattice reconstruction. As shown in Figs. \ref{stru_R}(a) and (b), the relaxed $R$-1.7$^{\circ}$ shows a pattern of triangular networks, consistent with previous studies and was also seen in the twisted bilayers of other TMD family members ~\cite{McGilly2020,Enaldiev2020,Weston2020,Rosenberger2020,Wijk2015,Yoo2019,Shi2020,VanWinkle2023,Tilak2023,Enaldiev2021,Rupp2023,Maity2021,Nieken2022,Huang2023}. The displacements of the reconstructed $R$-1.7$^{\circ}$ are shown in Figs.~\ref{stru_R}(c) and (d). The IP displacements of the Se atoms on the surface of MoSe$_2$ show a clockwise rotation around the RMX and RXM sites, while a counterclockwise rotation around the RMM site. As expected, the Se atomic displacements of WSe$_2$ are in opposite directions compared to those in MoSe$_2$. The OPP displacements for both the top and bottom layers show the same trend but exhibit different fluctuations, resulting in an apex at the RMM site and a basin at both the RMX and RXM sites. Figure~\ref{stru_R}(e) shows the interlayer distances for the relaxed $R$-1.7$^{\circ}$ moiré lattice, which exhibit the largest value at the RMM site. Figure~\ref{stru_R}(f) shows the changes in the reconstruction ($\delta_z$) and the layer distances for high-symmetry sites (RXM, RMX, and RMM) as the twist angle decreases. The behavior of $\delta_z$ exhibits a similar trend to that observed in the H-$\theta$ systems.
	
	\subsection{Electronic bands}
	\label{subsec:Electronic bands}
\begin{figure*}[!t]
		\includegraphics[width=1\linewidth]{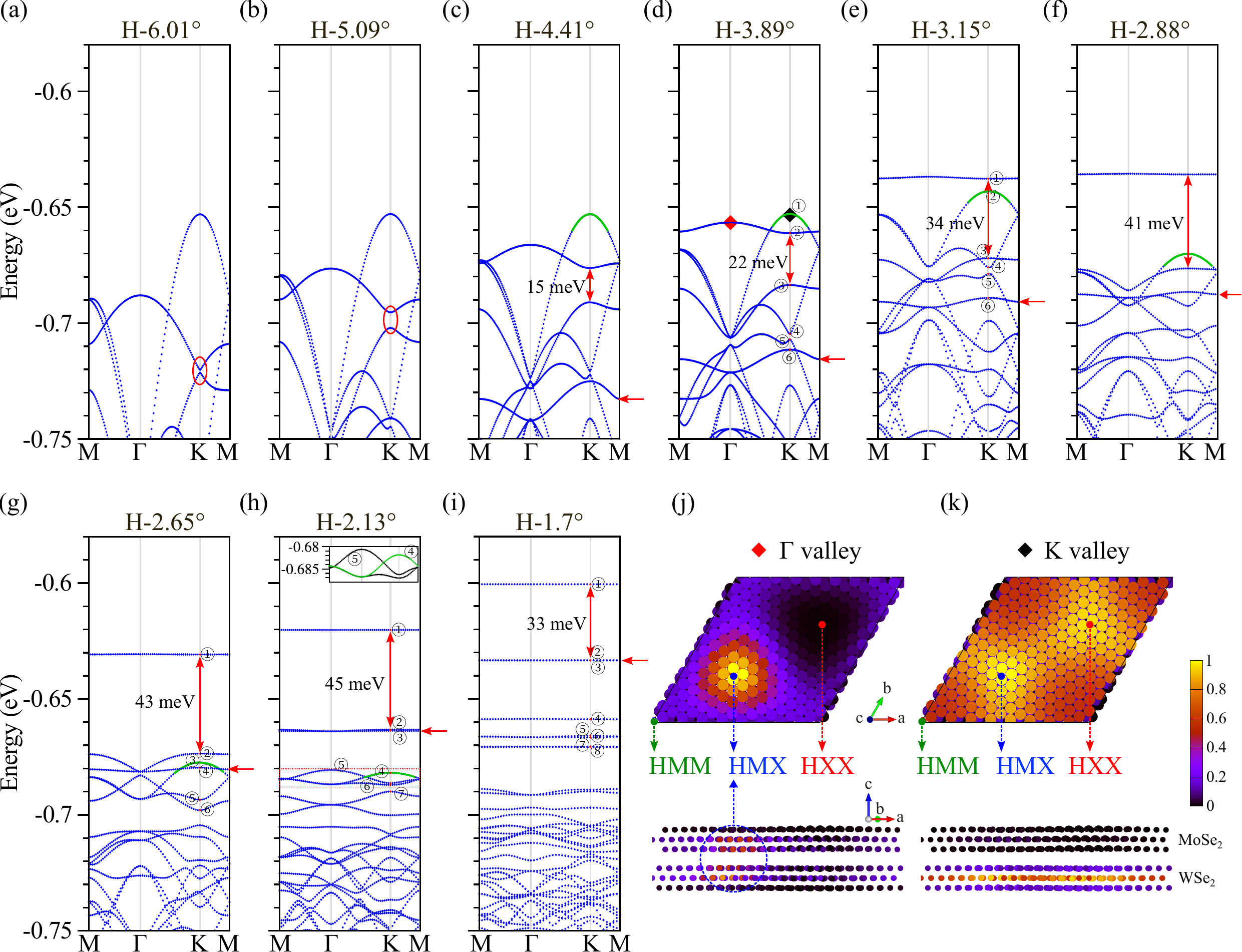}
		\caption{
			Electronic structures of twisted MoSe$_2$/WSe$_2$ heterobilayers. (a)-(i)~The valence bands of H-$\theta$. The Fermi level is shifted to the middle of the band gap. The red double-headed arrow marks the bandgap between the first and the second flat bands. The leftward arrow is used to track the evolution of band-6 of H-3.89$^{\circ}$ as the twist angle. The green trajectory is used to mark the K valley. (j, k)~The distribution of wave functions of the $\Gamma$ and K valleys (indicated by the red and black rectangles) for H-3.89$^{\circ}$ in (d). The upper panel and lower panel represent the top view and side view, respectively.
		}
		\label{band_H}
	\end{figure*}

	Figure~\ref{band_H} shows the band structure of MoSe$_2$/WSe$_2$ moiré lattices with the twist angle decreases from 6.01$^{\circ}$ to 1.7$^{\circ}$. Here, we focus on the valence bands, among which the highest six bands are labeled by numbers in Fig.~\ref{band_H}(d). One can see an almost dispersionless band (band-2) around -0.66 eV at the angle of 3.89$^{\circ}$, which is 3.7 meV below the valence band maximum at K (VBM$_K$). Its bandwidth is about 4.6 meV and reduces to only 0.7 meV at the angle of 3.15$^{\circ}$. Then, band-2 moves to be higher than VBM$_K$ and becomes increasingly flat. Note that band-3 and band-6 (guided by the red arrows) also move upward quickly as the twist angle decreases. As a result, there are three flat bands for $H$-2.13$^{\circ}$. Much more flat bands are obtained as the twist angle further decreases down to 1.7$^{\circ}$. In Fig.~\ref{band_H}(j, k), we analyze the wavefunction of the band at $\Gamma$ and K for $H$-3.89$^{\circ}$ [labeled by the diamonds in Fig.~\ref{band_H}(d)], respectively. One can see that the flat band at $\Gamma$ has contributions from both MoSe$_2$ and WSe$_2$ and is mainly localized at HMX. In contrast, the band at K is solely contributed by WSe$_2$. Orbital projections reveal that the flat band at $\Gamma$ is contributed by Se-$p_z$ and Mo/W-$d_{z^2}$ orbitals. Whereas the band at K is contributed by Se-$p_x$/$p_y$ and W-$d_{xy}/d_{x^2-y^2}$ orbitals (see Appendix~\ref{sec:orbital_projections} Fig.~\ref{band_H_21.79}). This situation is the same as the untwisted bilayer~\cite{Fang2015}. Therefore, the flat band (band-2) in Fig.~\ref{band_H}(d) is likely formed by the $\Gamma$ valley and the dispersive band (band-1) originates from the K valley, which is further supported by the unfolded bands discussed below. Moreover, one can also deduce that the $\Gamma$-valley flat band can be modulated by both the size of the moiré lattice and interlayer coupling. In contrast, one can expect that the interlayer coupling has a minor effect on the K-valley band, which forms a hexagonal lattice for $H$-3.89$^{\circ}$.  
	
	From Fig.~\ref{band_H}, one can also see how the flat bands evolve as the twist angle changes. In Fig.~\ref{band_H}(a, b), we marked a minigap at K by a circle. The formation of this gap is due to the interaction between the $\Gamma$ valley and its folded band. As the twist angle decreases, the BZ of the moiré lattice becomes smaller and smaller. Therefore, the $k$-point for the gap opening gets close to $\Gamma$. Meanwhile, the interlayer couplings are enhanced due to the reduced interlayer distance at HMX, which yields an increased minigap and pushes band-2 to high energies. These two facts lead to the quick flattening and a dramatic shift of the $\Gamma$-valley band. In contrast, the K-valley band (in green) is relatively less affected. Because it is modulated by the weak moiré potential, the gap openings at $\Gamma$ and M increase much slower than those for the $\Gamma$-valley flat band.

\begin{figure*}[ht]
		\includegraphics[width=1\linewidth]{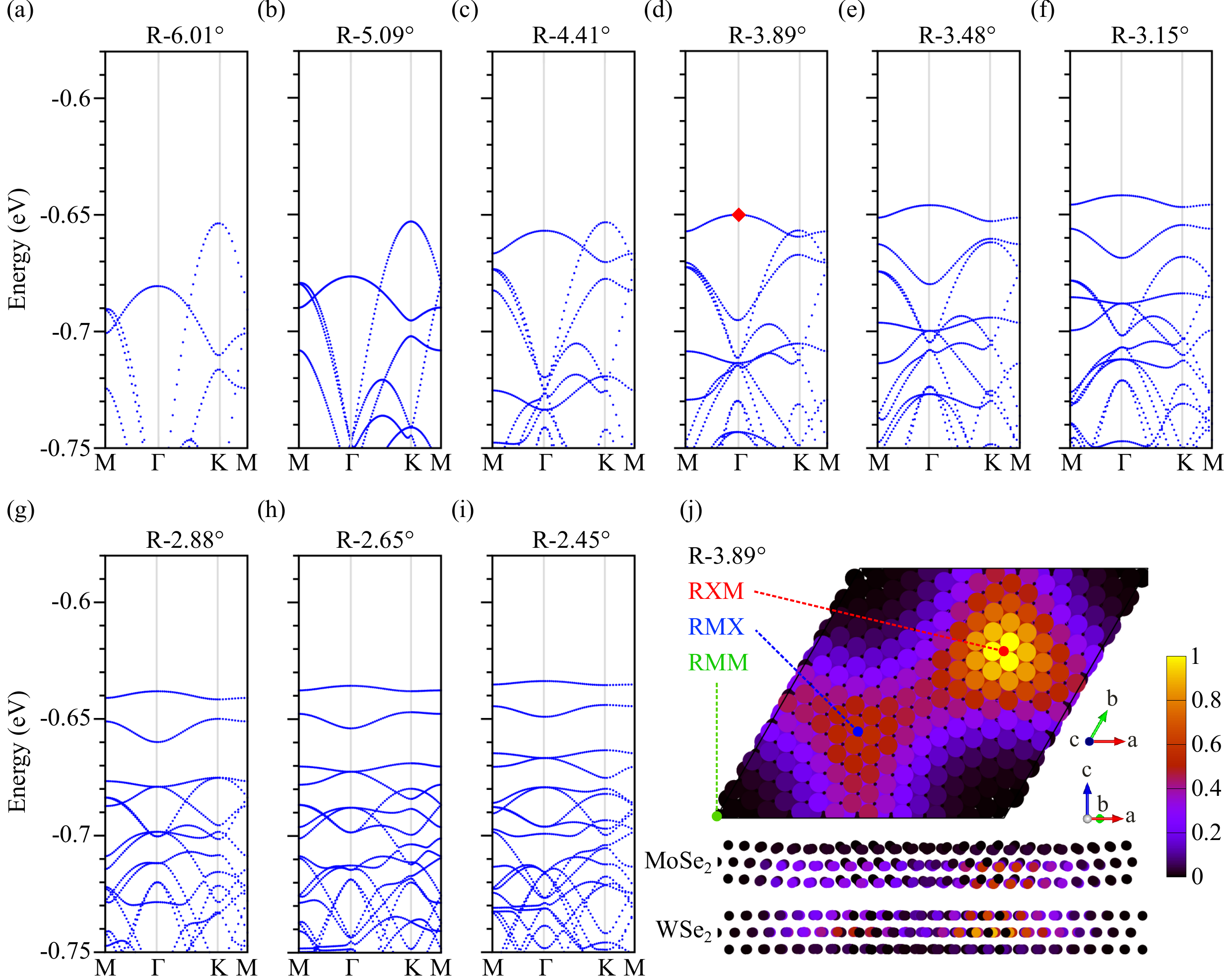}
		\caption{
			Band structure of R-stacked moiré lattices. (a)-(i)~The valence bands of $R$-$\theta$. (j)~Charge density distributions of the $\Gamma$ valley (guided by the red diamond) for $R$-3.89$^{\circ}$ in (d).
		} 
		\label{band_R}
	\end{figure*}

	Figure~\ref{band_R} shows the evolution of the flat bands in the R-stacking moiré lattice. As the angle decreases from 6.01$^{\circ}$ to 2.45$^{\circ}$, the $\Gamma$ valley gradually shifts to a higher energy level than VBM$_K$. A few flat bands emerge as the angle reaches 2.45$^{\circ}$. However, based on the analysis of the flat bands of the H-stacking moiré lattice, one can expect that these bands originate from the $\Gamma$-valley. We further analyzed the wavefunction of the $\Gamma$-valley flat bands of $R$-3.89$^{\circ}$ [see Fig.~\ref{band_R}(j)], which supports the above expectation. The wavefunction is mainly localized at RMX and RXM sites with interlayer hybridization. However, the RXM sites have contributed more than RMX due to symmetry breaking, which is different from the situation for the twisted homobilayers.
	
	There is a difference in the flat bands between the H- and R-stacking moiré structures. The bandwidth of the highest $\Gamma$-valley flat band of the R-stacking moiré patterns is larger than that of the H-stacking at the same angle. For example, the bandwidth of the highest $\Gamma$-valley flat band of R-3.89$^\circ$ is about 9.4 meV, which is double of that for H-3.89$^\circ$ [$\sim$4.6 meV, Fig.~\ref{band_H}(d)]. This difference in bandwidth may be related to the localization of high-symmetry stacking configurations. In the H-stacking moiré lattice, the wavefunction of the $\Gamma$-valley flat band is localized only at the HMX site [see Fig.~\ref{band_H}(j)]. Whereas the wavefunction of the highest flat band of the R-stacking moiré lattice is localized at both the RMX and RXM sites.

	\begin{figure*}[!t]
		\includegraphics[width=1\linewidth]{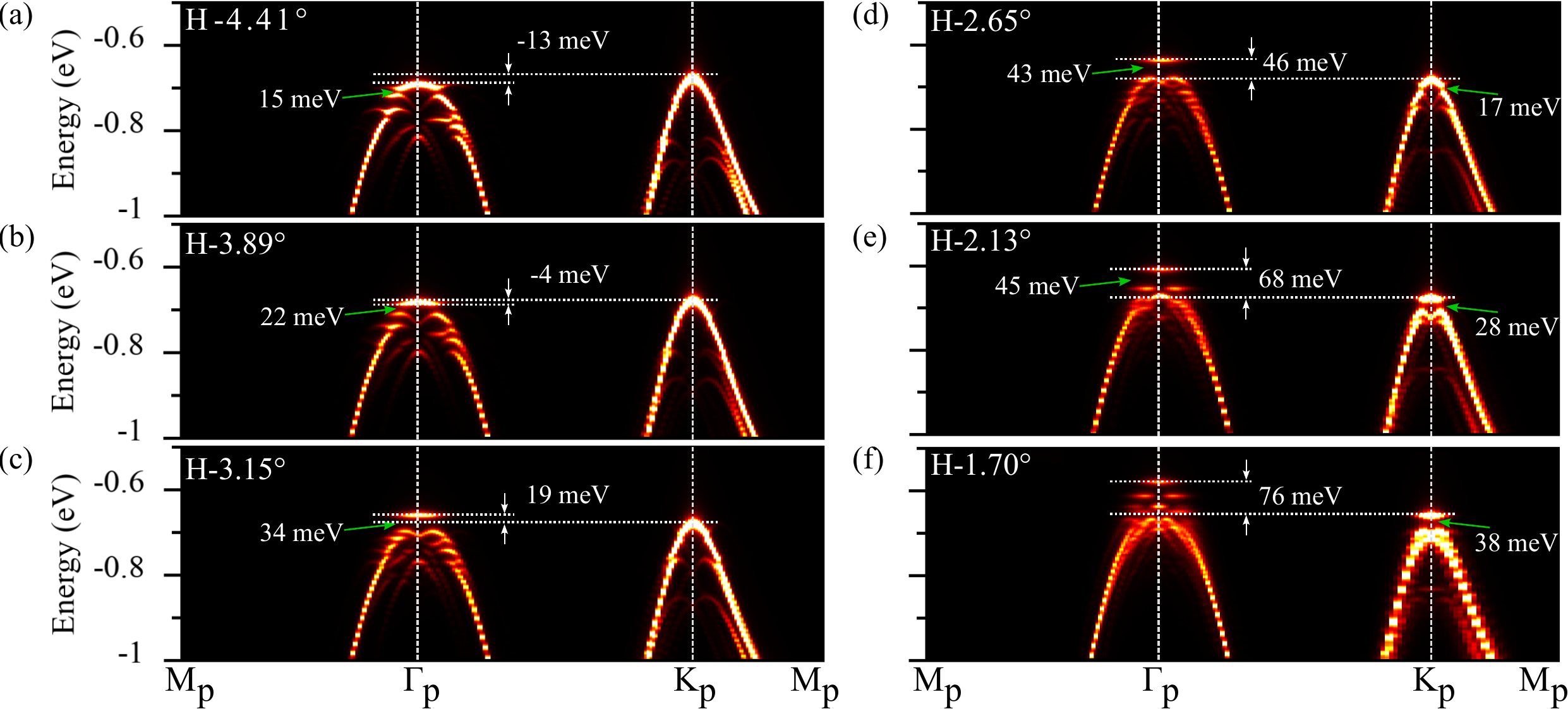}
		\caption{
			Unfoled bands for MoSe$_2$/WSe$_2$ moiré lattices near H-stacking. The sizes of the highest minigaps in both the $\Gamma$ and K valleys are shown and guided by the green arrows. The energy difference between the highest $\Gamma$-valley and K-valley flat bands is also shown (guided by the white arrows). 
		}
		\label{band_H_kproj}
	\end{figure*}
	
	We further obtained the unfolded band structure by projecting the wavefunctions onto the $k$-points in the BZ of the primitive cell of WSe$_2$, for which the results are shown in Fig.~\ref{band_H_kproj}. The reason is that WSe$_2$ contributes the highest K-valley flat band [see Fig.~\ref{band_H_21.79}(a) in Appendix~\ref{sec:orbital_projections}]. Indeed, one can see that for $H$-4.41$^{\circ}$ and $H$-3.89$^{\circ}$ the highest $\Gamma$-valley flat band is lower than VBM$_K$. Then it shifts to about 19 meV higher than VBM$_K$ for $H$-3.15$^{\circ}$. Therefore, one can again confirm that band-2 shown in Fig.~\ref{band_H}(d) originates from the $\Gamma$ valley. Likewise, by inspecting the energy differences between the bands at $\Gamma$ and K, one can also deduce that the three dispersionless bands (in numbers 1 - 3) shown in Fig.~\ref{band_H}(h) are $\Gamma$-valley flat bands. Moreover, an apparent K-valley flat band will not appear until the angle is decreased to 2.13$^{\circ}$, which is smaller than that for the $\Gamma$ valley.  This difference can be understood since the K-valley flat band is mainly caused by the structural reconstruction.

\begin{figure*}[!t]
		\includegraphics[width=1\linewidth]{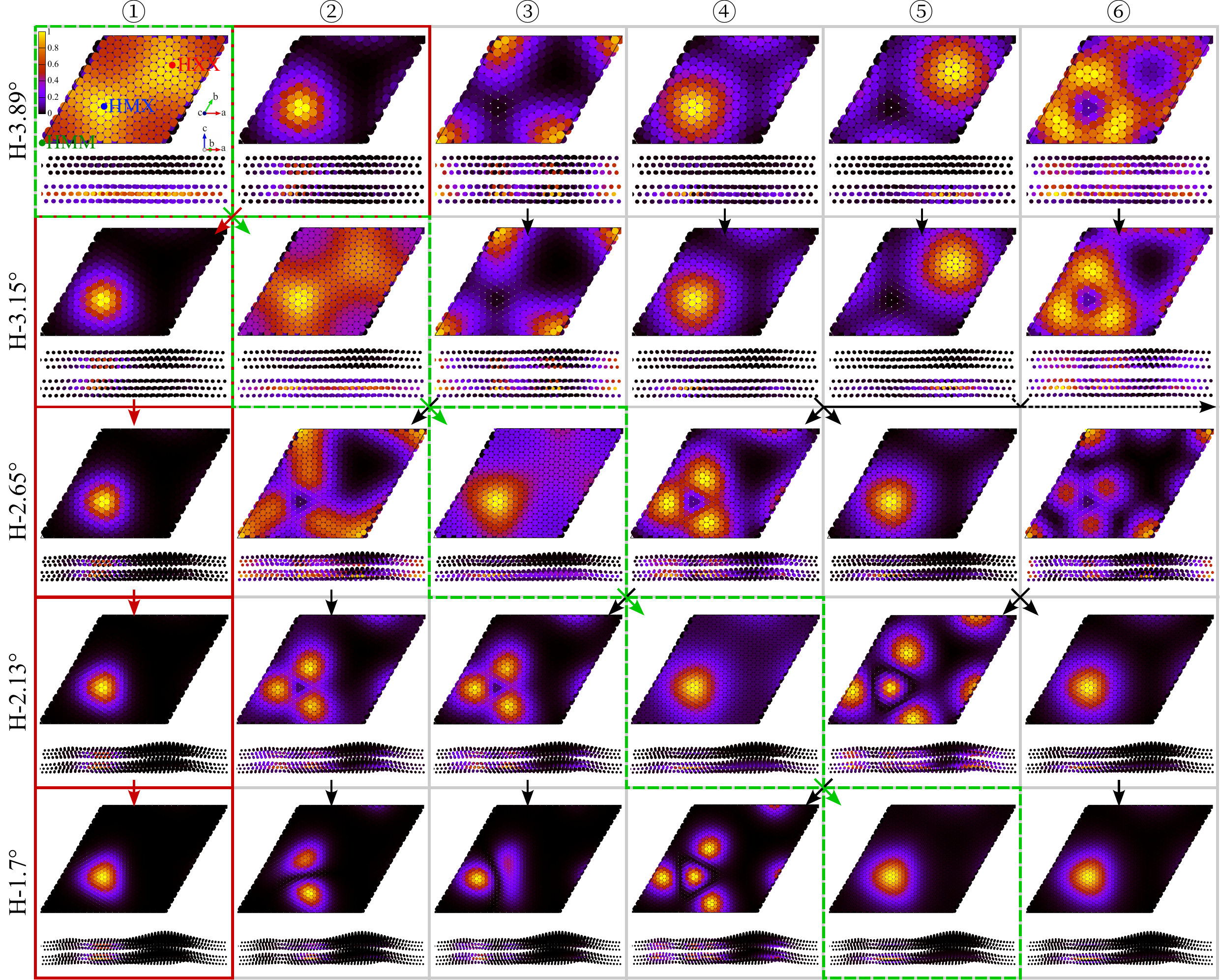}
		\caption{
			Wavefunctions of the six highest valence bands at K for twisted MoSe$_2$/WSe$_2$ heterobilayers in $H$-$\theta$ ($\theta$ = 3.89$^{\circ}$, 3.15$^{\circ}$, 2.65$^{\circ}$, 2.13$^{\circ}$, and 1.7$^{\circ}$). The numbers represent the bands shown in Fig.~\ref{band_H}. The arrows are used to guide the eye in seeing the evolution of the electronic states. The green arrow is for the K-valley flat band and the red one is for the highest $\Gamma$-valley flat band. The black ones are for the rest. In each panel, both top and side views are given.  
		}
		\label{wavefunction_H}
	\end{figure*}

	Figure~\ref{wavefunction_H} shows the evolution of wavefunctions of the flat bands at K for the MoSe$_2$/WSe$_2$ moiré lattices as a function of the twist angle. For $H$-3.89$^\circ$, the VBM$_K$ shows a hexagonal lattice. As the twist angle decreases, the weight of HMX gets larger and larger. At 2.13$^\circ$, it becomes completely localized at HMX, which is consistent with the situation that the K-valley flat band emerges at this angle. Consequently, a transition from the hexagonal lattice to the triangular lattice occurs for the K-valley band. At $\theta=3.89^\circ$, the $\Gamma$-valley flat band (band-2) exhibits a triangular lattice. Although band-2 shifts to higher energy as the twist angle decreases, its profile remains unchanged. Similarly, the profile of band-3 is also preserved. However, band-3 experiences significant changes, which evolves into a three-lobe pattern as the twist angle is decreased to 2.13$^\circ$. Bands 4 and 5 exhibit a triangular lattice and the characteristics of K valley since the wavefunctions are entirely localized at WSe$_2$. Band-6 also exhibits a hexagonal lattice, which evolves into a three-lobe pattern at $\theta$=2.65$^\circ$ (now denoted as band-4) and 2.13$^\circ$ (band-3). At this angle bands 2 and 3 are almost energetically degenerate [see Fig.~\ref{band_H}(h)]. Therefore, they interact under the effect of moiré potential. As a result, the patterns of the wavefunctions also experience changes (see bands 2 and 3 at $\theta$=1.7$^\circ$). 
	
	\subsection{Effect of spin-orbit coupling}
	\label{subsec:Effect of spin-orbit coupling}
\begin{figure}[!t]
		\includegraphics[width=1\linewidth]{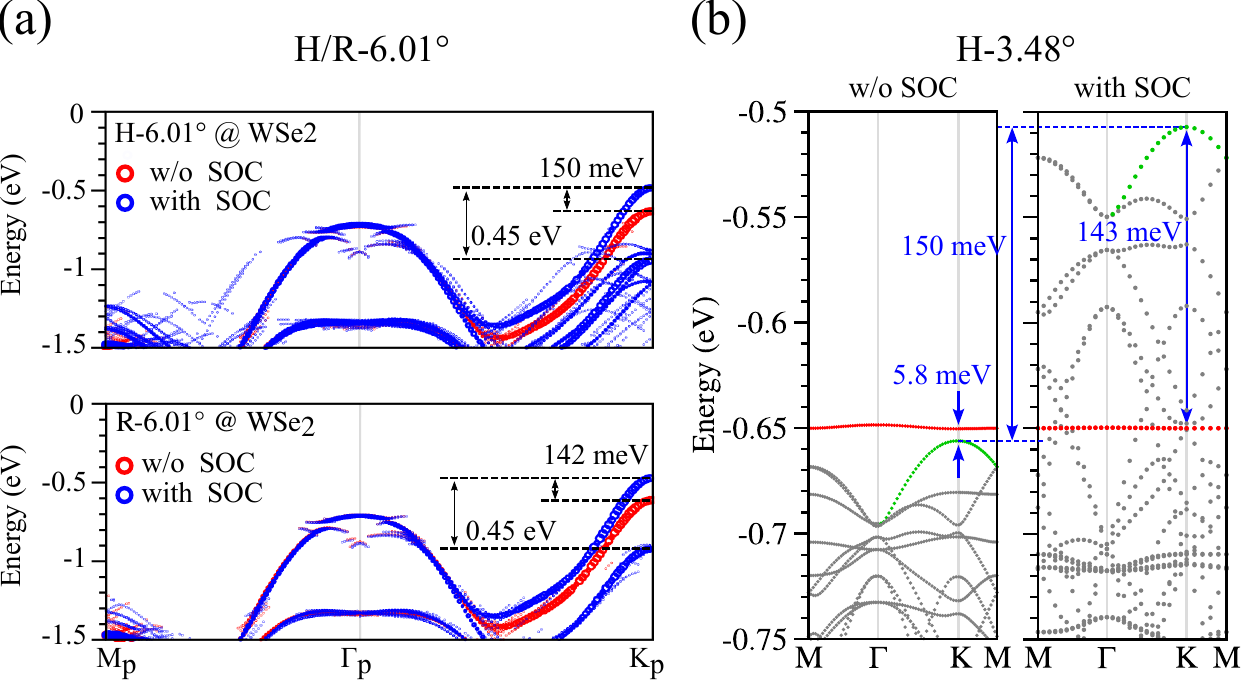}
		\caption{
				Effect of SOC on the band structures of 6.01$^\circ$ and 3.48$^\circ$ moiré lattices.
				(a) $k$-projected bands with (blue) and without (red) SOC for WSe$_2$ in $H$-6.01$^{\circ}$ and $R$-6.01$^{\circ}$.
				(b) The band structure of $H$-3.48$^\circ$ with band folding. The $\Gamma$-valley flat bands (in red) are placed at the same energy for comparison. The K-valley bands are shown in green.
		}
		\label{band_soc}
	\end{figure}

	We now discuss the effect of SOC on the flat bands. Figure~\ref{band_soc}(a) shows the unfolded bands with layer projections onto WSe$_2$ with and without SOC for $H$-6.01$^{\circ}$ and $R$-6.01$^{\circ}$. One can see that the bands near $\Gamma$ remain unchanged with the inclusion of SOC. In contrast, there is a SOC splitting of about 0.45 eV at K for both the $H$-6.01$^{\circ}$ and $R$-6.01$^{\circ}$ systems. Such a splitting is very close to that for the untwisted MoSe$_2$/WSe$_2$ heterobilayers [Fig.~\ref{layer_dis_band_H_R}(b) in Appendix~\ref{sec:untwisted_bilayer}]. Note that the $\Gamma$-valley flat band is only 76 meV higher than the K-valley flat band without SOC [Fig.~\ref{band_H_kproj}(f)]. Therefore, the K-valley band should be the VBM for MoSe$_2$/WSe$_2$ moiré lattices.  We further investigate the SOC effect on the band structure of H-3.48$^\circ$, for which the K valley is lower than the $\Gamma$-valley flat band by about 5.8 meV without SOC [Fig.~\ref{band_soc}(b)]. The inclusion of SOC leads to that the K valley is about 143 meV higher than the $\Gamma$-valley flat band. Moreover, our results also show that the SOC splitting has a weak dependence on twisting and stacking.
	
	\section{CONCLUSIONS}
	\label{sec:CONCLUSIONS}
	In conclusion, we have investigated the electronic structure of MoSe$_2$/WSe$_2$ moiré lattices by combining machine-learning methods and a band unfolding technique, which enables tracking the evolution of flat bands as a function of the twist angle. We find that multiple flat bands emerge when the twist angle is decreased to 2.13$^{\circ}$. We reveal that the interlayer coupling and moiré potential are responsible for the emergence of the $\Gamma$-valley flat bands.  Whereas, the K-valley flat band is mainly modulated by the structural reconstruction.  Due to this difference, the twist angle for generating the $\Gamma$-valley flat bands is larger than that for the K-valley. And the $\Gamma$-valley flat bands shift to higher energies much faster than the K-valley flat band. By analyzing the characteristics of the wavefunctions of the valence bands at K, we find that the K-valley flat band experiences a triangular-to-hexagonal lattice transition. We have further discussed the effect of SOC on the flat bands of the twisted systems. Our results find that like the untwisted TMD bilayers, the SOC has a minor effect on the $\Gamma$-valley flat bands but induces a significant splitting for the K-valley. By including the SOC, the K-valley flat bands are higher in energy than the $\Gamma$-valley flat bands, that is, the K-valley flat bands contribute to the VBM of the MoSe$_2$/WSe$_2$ moiré lattices.
	
	\section*{ACKNOWLEDGMENTS}
	This work was supported by the National Natural Science Foundation of China (Grants No. 12174098 and No. 11774084) and by the State Key Laboratory of Powder Metallurgy, Central South University, Changsha, China. Calculations were carried out in part using computing resources at the High-Performance Computing Platform of Hunan Normal University.

	\appendix

        \section{DEEPH MODEL}
	\label{sec:deeph model}
\begin{figure}[!t]
		      \includegraphics[width=1\linewidth]{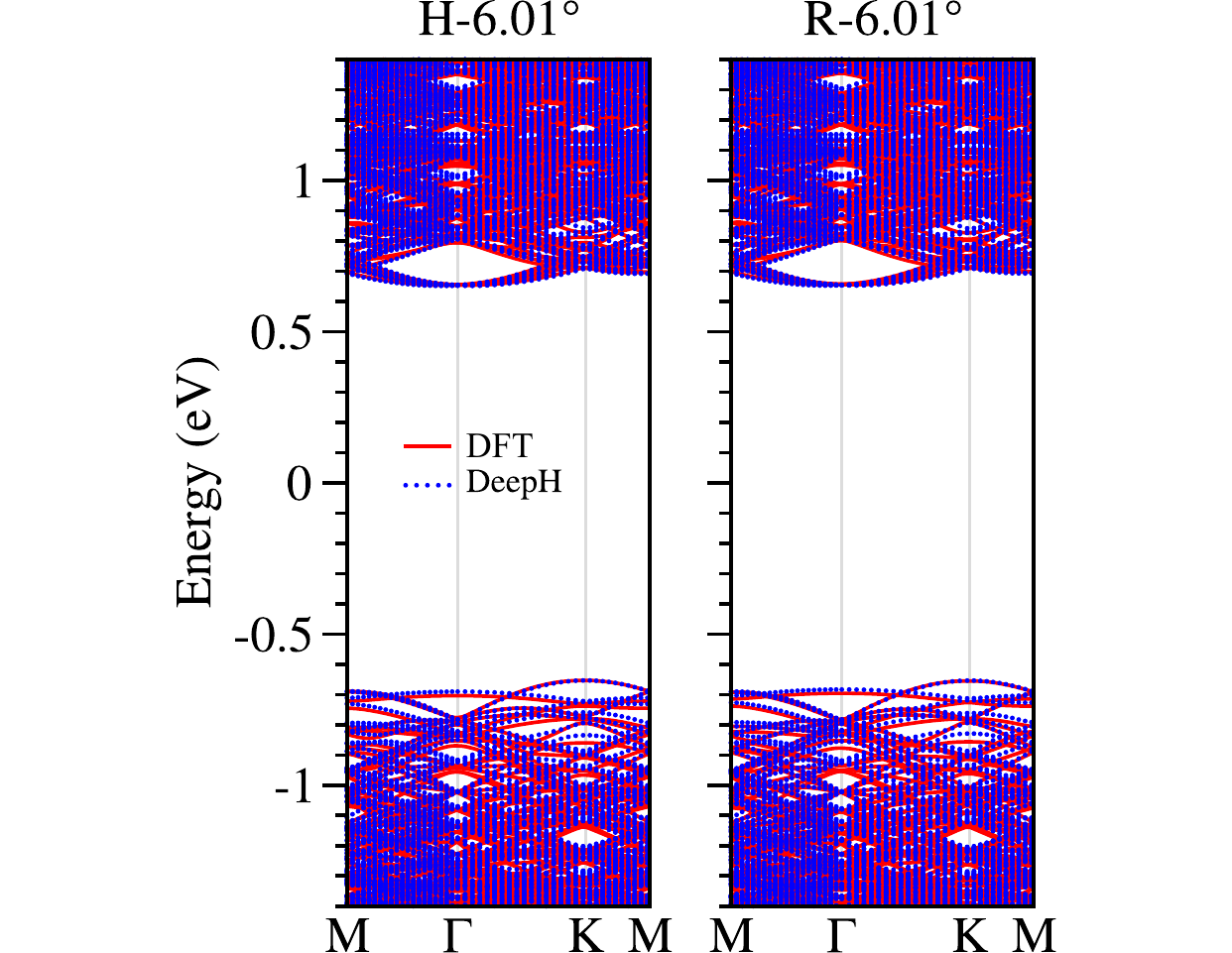}
		      \caption{
Benchmark DeepH calculations of $H$- and $R$-6.01$^\circ$. DFT results are also shown for comparison.}
		      \label{band_abacus_deeph_H_R_6.01}
	    \end{figure}

        Figure~\ref{band_abacus_deeph_H_R_6.01} shows the band structures of $H$-6.01$^\circ$ and $R$-6.01$^\circ$. It should be mentioned that our dataset does not include any moiré structures with the twist angle of 6.01$^\circ$. The band structures obtained by the DeepH model agree well with those from DFT, confirming the high precision of our DeepH model.
 
	\section{UNTWISTED BILAYER}
	\label{sec:untwisted_bilayer}
\begin{figure}[!t]
		\includegraphics[width=1\linewidth]{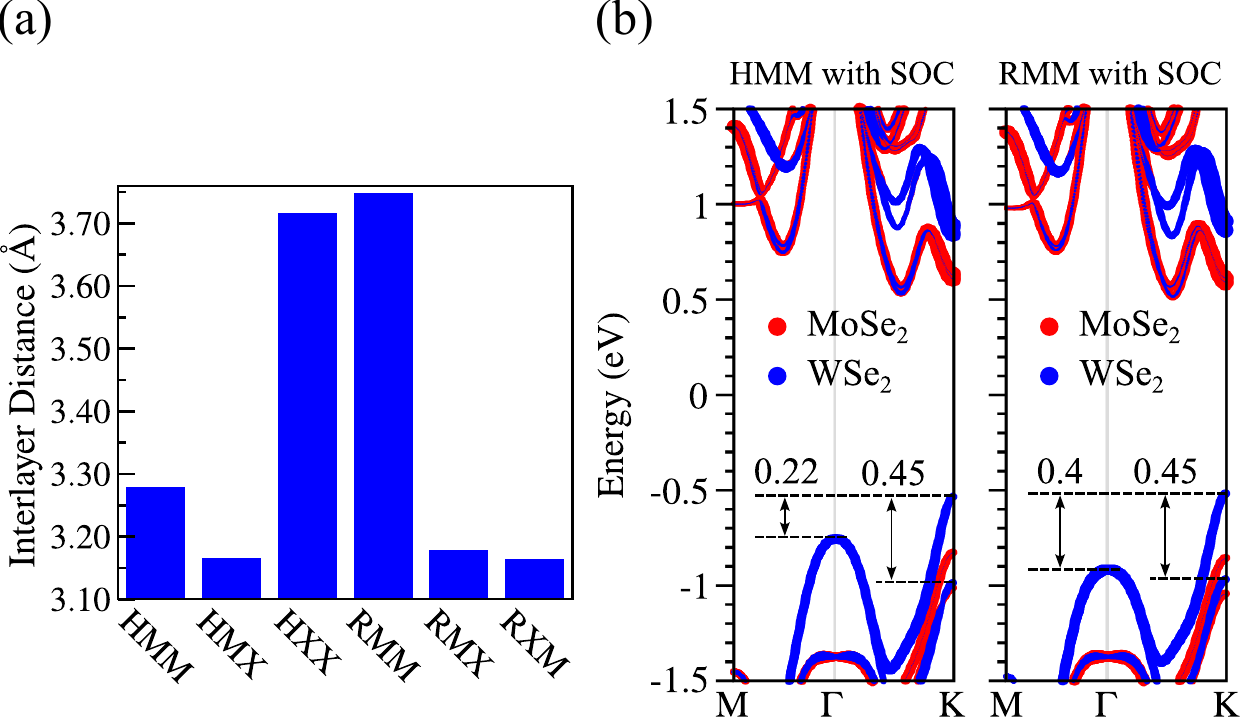}
		\caption{
			Interlayer distances and band structures of the untwisted bilayers.
			(a)~Interlayer distances of six high-symmetry configurations for the H-stacking and R-stacking.
			(b)~Band structures of HMM and RMM with SOC. Red and blue represent MoSe$_2$ and WSe$_2$, respectively. The energy difference between the highest $\Gamma$ and K valleys, as well as the SOC splitting of the K valleys, are shown by the black arrows.
		} 
		\label{layer_dis_band_H_R}
	\end{figure}

	Figure~\ref{layer_dis_band_H_R}(a) shows the interlayer distances of untwisted bilayer systems for H- and R-stacking. For H-stacking, the HXX configuration has the largest interlayer distance, while HMX has the smallest. Similarly, RMM exhibits the largest interlayer distance for R-stacking, and RXM the smallest. We further calculated the band structures for HMM and RMM, which are shown in Fig.~\ref{layer_dis_band_H_R}(b). The band splitting at the K valley, induced by SOC for both stackings, is 0.45 eV. Additionally, the K valley of HMM is 0.22 eV higher than the $\Gamma$ valley. For RMM, the energy difference between the two valleys is 0.4 eV.
	
	\section{ORBITAL PROJECTIONS}
	\label{sec:orbital_projections}
\begin{figure}[!t]
		\includegraphics[width=1\linewidth]{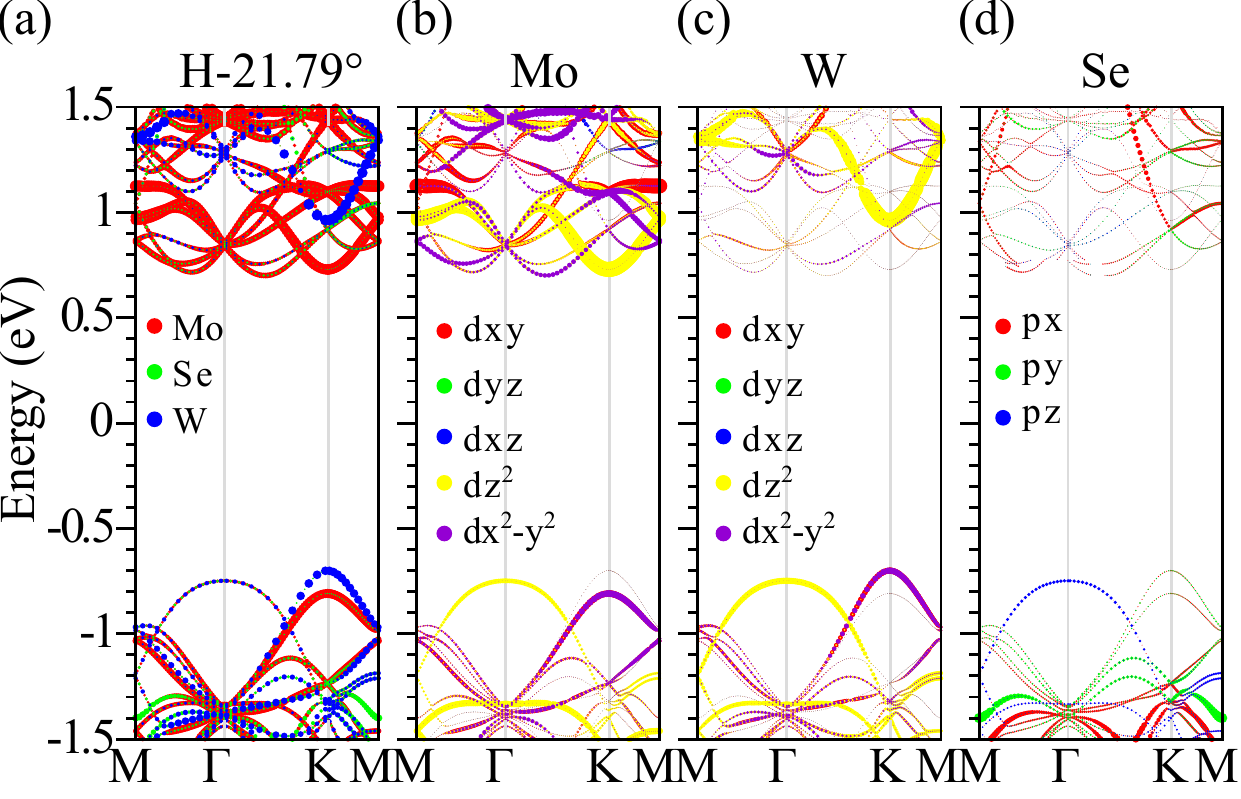}
		\caption{
			Band structures of $H$-21.79$^{\circ}$.
			(a)~Band structure projected onto Mo (red), W (blue), and Se (green).
			(b)-(d)~Band structure projected onto the orbitals of Mo, W, and Se, respectively.
		} 
		\label{band_H_21.79}
	\end{figure}

	Figure~\ref{band_H_21.79} shows the band structures of $H$-21.79$^{\circ}$ with layer- and orbital-projections. One can see that the $\Gamma$ valley is contributed by both MoSe$_2$ and WSe$_2$, and the highest K valley is contributed solely by WSe$_2$. Furthermore, the $\Gamma$ valley is dominated by Mo/W-$d_{z^{2}}$ and Se-$p_z$ orbitals, whereas the highest K valley is mainly contributed by W-$d_{xy}/d_{x^2-y^2}$ and Se-$p_z$ orbitals. Consequently, the atomic orbitals contributed to the $\Gamma$ valley are all oriented in the OOP direction, resulting in strong interlayer hybridization that pushes the band up to higher energies as the interlayer distance decreases. In contrast, the orbitals in the K valley are oriented in the IP direction, to which the interlayer hybridization has a minor effect.
	
	\bibliographystyle{apsrev4-2}
	\bibliography{references}
	
\end{document}